\documentclass[english,tightenlines,eqsecnum,amsmath,amssymb,aps,prd,superscriptaddress,nofootinbib,showpacs,floats]{article}
\usepackage{amsmath,amsfonts,amssymb,multicol}
\usepackage{graphicx,caption,subcaption}
\usepackage{epstopdf}
\usepackage[usenames]{xcolor}
\usepackage{epstopdf}
\definecolor{AHZ}{rgb}{0.0,0.9,0.2}
\usepackage[colorlinks,linkcolor=AHZ,citecolor=red,bookmarks]{hyperref}
\usepackage{rotating}
\usepackage[mathscr]{eucal}
\usepackage{graphicx}
\usepackage[T1]{fontenc}
\usepackage{babel}
\usepackage{authblk}
\usepackage{epsfig}
\usepackage{color}
\usepackage{amssymb}
\usepackage{fancyhdr}

\def\be{\begin{equation}}
\def\ee{\end{equation}}
\def\bea{\begin{eqnarray}}
\def\eea{\end{eqnarray}}

\def\bwt{\begin{widetext}}
	\def\ewt{\end{widetext}}

\begin{document}
%
%
\title{Black hole solutions and Euler equation in Rastall and generalized Rastall theories of gravity}
\author{H. Moradpour$^1$\footnote{Corresponding author: h.moradpour@riaam.ac.ir}, Y. Heydarzade$^{2}$\footnote{yheydarzade@bilkent.edu.tr}, C. Corda$^{3}$\footnote{cordac.galilei@gmail.com},
A. H. Ziaie$^{1}$\footnote{ah.ziaie@riaam.ac.ir}, S.
Ghaffari$^{1}$\footnote{sh.ghaffari@riaam.ac.ir}}
\affil{$^{1}$ Research Institute for Astronomy and Astrophysics of Maragha (RIAAM), P.O. Box
	55134-441, Maragha, Iran\\
$^2$ Department of Mathematics, Faculty of Sciences, Bilkent University, 06800 Ankara, Turkey\\
$^3$ International Institute for Applicable Mathematics and
Information Sciences, B. M. Birla Science Centre, Adarshnagar,
Hyderabad 500063, India}

\maketitle


\begin{abstract}
Focusing on the special case of generalized Rastall theory, as a
subclass of the non-minimal curvature-matter coupling theories in
which the field equations are mathematically similar to the
Einstein field equations in the presence of cosmological constant,
we find two classes of black hole (BH) solutions including $i$)
conformally flat solutions and $ii$) non-singular BHs. Accepting
the mass function definition and by using the entropy contents of
the solutions along with thermodynamic definitions of temperature
and pressure, we study the validity of Euler equation on the
corresponding horizons. Our results show that the thermodynamic
pressure, meeting the Euler equation, is not always equal to the
pressure components appeared in the gravitational field equations
and satisfies the first law of thermodynamics, a result which in
fact depends on the presumed energy definition. The requirements
of having solutions with equal thermodynamic and Hawking
temperatures are also studied. Additionally, we study the
conformally flat BHs in the Rastall framework. The consequences of
employing generalized Misner-Sharp mass in studying the
validity of the Euler equation are also addressed.\\


\noindent Pacs: 04.50.Kd; 04.70.-s
\end{abstract}

\maketitle

\section{Introduction\label{Intr}}

The conservation of energy-momentum forms the backbone of the
general relativity theory (GR) \cite{pois,gelen}. Its breakdown
also helps us in providing an explanation for observations,
confirming the current acceleration of the universe \cite{prln}.
Indeed, GR is modified by admitting a non-minimal coupling between
geometry and matter fields
\cite{rastall,cmc,cmc1,cmc2,genras,rl3,rl4,rl5,rl1,rl2}.
Theoretically, the origin of these attempts comes back to P.
Rastall \cite{rastall} who argued that the ordinary
energy-momentum conservation law (OCL) can be violated in the
curved spacetime, a hypothesis which leads to interesting
cosmological and gravitational consequences \cite{more}. Even
whenever OCL is valid, and the non-minimal coupling has not
happened, the existence of such ability in the structure of
geometry and matter fields can provide proper description for the
cosmic eras \cite{genras}.

Black hole solutions have a lot of significance from both
theoretical and experimental perspectives \cite{pois,gelen}. For
example, one can attribute the Hawking temperature to their
horizon, and use the Misner-Sharp mass to write the gravitational
field equations as a thermodynamic equation of state
\cite{11,ms,mv}. Therefore, in this approach, the energy
definition has a crucial role \cite{11,ms,mv}, a role which shows
again itself in modeling the cosmic evolution \cite{mpla}. It is
also useful to mention that the Hawking temperature is not always
equal to the thermodynamic temperature obtained from the
thermodynamic definition of temperature (see Ref. \cite{thermotbh}
and references there in), and indeed, the deep connection between
gravity and thermodynamics needs to be further studied
\cite{11,thermotbh}.

For a thermodynamic system with entropy $S$, energy $E$, pressure
$P$ and temperature $T$ within the volume $V$, the Euler equation
takes the form $E=TS-PV$ \cite{callen}. The quality of validity of
the Euler equation for BH solutions of various gravitational
theories have not yet been fully studied \cite{11,ms,mv}. In fact,
the study of the quality of validity of the Euler equation can
help us in achieving a better understanding of relation between
thermodynamics and gravity, and also the concept of the quantities
such as energy.

Although most of the known BH solutions are singular at their
spatial origin, i.e $r=0$, it has been shown that there are also
non-singular BH solutions in the framework of GR \cite{non}. On
the other hand, conformal flat spacetimes, for which the Weyl
tensor is zero, has interesting properties \cite{pois,gelen}
motivating physicists to study them \cite{confbh}. Recently, some
BH solutions have been derived in the Rastall theory and its
generalized version \cite{bhras,nonras1,nonras,rasbh} showing that
the de Sitter spacetime, a conformal flat and non-singular
spacetime, can be obtained as a vacuum solution in these theories
\cite{bhras,rasbh}.

All of the above arguments motivate us to look for the conformally
flat and non-singular BHs in the context of Rastall theory as well
as in a special subclass of the generalized Rastall theory, which
its field equations are similar to the Einstein equations with the
cosmological constant. After addressing the field equations
corresponding to the generalized Rastall theory and also the
Rastall theory in section 2, we obtain conformally flat BH
solutions and investiagte the validity of the Euler equation in
both mentioned frameworks in the third section. In the section 4,
we discuss non-singular BH solutions in the framework of a special
subclass of the generalized Rastall theory. In section 5, we
present our conclusion. Throughout this work, we use the units so
that $c=\hbar=k_B=1$, where $k_B$ denotes the Boltzmann constant.

\section{Field equations}

In the generalized Rastall theory \cite{genras}, OCL is modified
as

\begin{eqnarray}\label{gr0}
T^{\mu\nu}_{\ \ \ ;\mu}=(\lambda R)^{;\nu},
\end{eqnarray}

\noindent leading to the field equations

\begin{eqnarray}\label{gr}
G_{\mu\nu}+\kappa\lambda g_{\mu\nu}R=\kappa T_{\mu \nu},
\end{eqnarray}

\noindent where $\kappa$ and $\lambda$ denote the Rastall
gravitational coupling constant and the Rastall parameter,
respectively. Clearly, Eq.~(\ref{gr0}) indicates the validity of
OCL does not necessarily mean that $\lambda=0$ (OCL is met
whenever $\lambda=\frac{\beta}{R}$ where $\beta$ is an unknown
constant). For this theory, the (anti)de Sitter spacetime can be
obtained as the vacuum solution \cite{genras}. Now, let us
consider the static spherically symmetric metric

\begin{equation}\label{met}
ds^2=-f(r)dt^2+f^{-1}(r)dr^2+r^2d\Omega_2^2,
\end{equation}

\noindent filled by a source satisfying OCL. In this manner, the
field equations~(\ref{gr}) take the form

\begin{eqnarray}\label{gr1}
G_{\mu\nu}=\kappa T_{\mu \nu}-\kappa\beta g_{\mu\nu}=\kappa(T_{\mu
\nu}-\beta g_{\mu\nu}),
\end{eqnarray}

\noindent which obviously confirms that the (anti)-de Sitter
spacetime is a vacuum solution meaning that the $\kappa\beta$ term
plays the role of cosmological constant \cite{genras}.

Mathematically, the field equations (4) and the Einstein field
equations in the presence of cosmological constant (EECC) have
similar forms. In fact, for EECC we have $G_{\mu\nu}=\kappa_E
T_{\mu \nu}-\Lambda g_{\mu\nu}=\kappa_E(T_{\mu
\nu}-\frac{\Lambda}{\kappa_E} g_{\mu\nu})$, where $\kappa_E$ is
the Einstein gravitational coupling constant. Here, $\Lambda$
represents the cosmological constant and $\frac{\Lambda}{\kappa_E}
g_{\mu\nu}$ denotes its corresponding energy-momentum tensor.
Comparing with Eq.~(4), we find that the $\kappa\beta$ term is the
counterpart of $\Lambda$ in this formalism. Hence, we have
$\rho=-p=\beta$ for the energy density ($\rho$) and pressure ($p$)
of this source. However, there are  four differences between
Eq.~(4) and EECC as $i$) in Eq.~(4), the $\kappa\beta$ term,
playing the role of cosmological constant, is naturally arising
from the ability of spacetime and matter fields to couple with
each other in a non-minimal way (i.e.  $\beta\neq0$), and it has
not manually been added to the field equations as in GR, $ii$) the
naturally arising cosmological constant of Rastall theory
($\kappa\beta$) is proportional to the gravitational coupling
$\kappa$, a feature which does not exist in GR at least at the
level of field equations, $iii$) the Rastall gravitational
coupling constant $\kappa$ is not necessarily equal to that of the
Einstein theory unless one sets $\lambda=0$, and $iv$) as we will
show, the Misner-Sharp content of mass is different in the
frameworks of EECC and Rastall theory. These differences lead to
different thermodynamic features for the obtained solutions which
are addressed in the paper.

Whenever $\lambda=constant\equiv\xi$, Eq.~(\ref{gr0}) reduces to
the original Rastall hypothesis \cite{rastall} in which

\begin{eqnarray}\label{r1}
T^{\mu\nu}_{\ \ \ ;\mu}=\xi R^{,\nu}~~\Rightarrow~~
G_{\mu\nu}+\kappa_R\xi g_{\mu\nu}R=\kappa_R T_{\mu \nu},
\end{eqnarray}

\noindent where $\kappa_R$ denotes the original Rastall
gravitational coupling which generally differs from $\kappa$ and
that of the Einstein theory \cite{rastall,genras}.

Investigation of the Misner-Sharp mass in the Rastall and
generalized Rastall frameworks, as well as the Hawking
temperature, horizon entropy and the validity of the first law of
thermodynamics on the horizons of metric~(\ref{met}) have been
studied in Refs. \cite{ms,mv}. For this propose, the Clausius
relation, the first law of thermodynamics together with the $t-t$
and $r-r$ components of the field equations are used
\cite{ms,mv,11}.

\section{Conformally flat Black hole solutions}
\subsection{Conformally flat BHs in the generalized Rastall theory}

In order to find the conformal flat solutions of~(\ref{gr1}), the
Weyl tensor defined as

\begin{eqnarray}\label{weyl}
&&C_{ijkl}=R_{ijkl}-\frac{1}{n-1}(g_{ijkp}R^p_{\ l}+g_{ijpl}R^p_{\
k})\nonumber\\&&+\frac{1}{n(n-1)}g_{ijkl}R,
\end{eqnarray}

\noindent should be zero \cite{pois}. This gives the solution

\begin{equation}\label{met1}
f(r)=1-ar-br^2,
\end{equation}

\noindent where $a$ and $b$ are arbitrary integration constants.
Setting $a = 0$ and $b = \Lambda$, the line element represents the
de Sitter spacetime. Then, this solution is a generalization to de
Sitter solution and is also similar to the solution introduced by
Kiselev, except possessing of a mass term  \cite{kis}. The $a$
parameter in (\ref{met1}) plays the role of a quintessence
field in \cite{kis}. Then, depending on the $a$ and $b$ parameter
values, our metric function can represent a flat, quintessence and
de Sitter spacetime, respectively. One may also consider  the
limiting behavior of this solution for small and large $r$ values
where each of the quintessence  and cosmological constant fields
dominate.

Now, using the field equations~(\ref{gr}), one can easily reach at

\begin{equation}\label{met2}
b=\frac{\kappa\beta}{3},~~
\rho(r)=-p_r(r)=-2p_t(r)=\frac{2a}{\kappa r},
\end{equation}

\noindent as a solution in which $p_r(r)$ and $p_t(r)$ are the
radial and transverse pressure components, respectively. We
clearly see that the obtained solution~(\ref{met2}) addresses an
anisotropic fluid. As usual, $\rho(r)$ denotes the energy density,
and respecting the weak energy condition requires to have
$a\kappa>0$. From the energy-momentum tensor components in
(\ref{met2}), one also realizes that the quintessence field here
has an average equation of state of $\omega_{av}=-\frac{2}{3}$,
indicating a gravitationally repulsive fluid. Although the metric
with (\ref{met1}) is well-behaved at the center ($r=0$), the
components of the corresponding $T_{\mu \nu}$ in (\ref{met2})
diverge at the $r\rightarrow0$ limit, which signals a physical
singularity. In fact, unlike the Weyl square, the Kretschmann
invariant, the Ricci scalar and the Ricci square diverge at this
limit. The $b=\frac{\kappa\beta}{3}$ relation is also interesting,
because it is claiming that the value of the cosmological constant
specifies $b$, meaning that the value of $\beta$ affects the
location of the spacetime horizons (the solutions of the
$f(r_h)=0$ equation). In fact, the (anti)-de Sitter sector of
metric ($b\neq0$) is cancelled in the field equations~(\ref{gr1})
by adopting the special value of $\beta$ (and thus $\lambda$)
satisfying the $b=\frac{\kappa\beta}{3}$ condition. For this
solution, whenever $f(r)=1-ar$, OCL is met, and the
energy-momentum tensor obtained in Eq.~(\ref{met2}), satisfies the
$G_{\mu\nu}=\kappa T_{\mu \nu}$ equations.

Here, instead of the generalized Misner-Sharp mass \cite{mv}, we
use the mass function corresponding to the energy density $\rho$
defined as \cite{non}

\begin{eqnarray}\label{mass0}
m(r)=4\pi\int_{0}^{r}\rho(r)r^2dr,
\end{eqnarray}

\noindent and can be combined with the obtained energy
density~(\ref{met2}) to reach \cite{non}

\begin{eqnarray}\label{mass1}
m(r)=4\pi\int_{0}^{r}\rho(r)r^2dr=\frac{4\pi a}{\kappa}r^2,
\end{eqnarray}

\noindent leading to

\begin{eqnarray}\label{mass20}
m(r_h)\equiv M=\frac{a}{\kappa}A,
\end{eqnarray}

\noindent for the mass circumscribed by the horizon located at
$r_h$ with area $A=4\pi r_h^2$. In the generalized Rastall theory,
the entropy ($S$)-area ($A$) relation is written as
$S=\frac{2\pi}{\kappa}A$ \cite{mv} combined with Eq.~(\ref{mass1})
to reach at $M=\frac{a}{2\pi}S$. Now, the thermodynamic
temperature definition ($T=\frac{\partial M}{\partial S}$) yields
$T=\frac{a}{2\pi}$ for the temperature of the energy source
circumscribed by the radius $r_h$. Then, from the weak energy
condition and the positivity of the entropy $S$, one finds that
both $a$ and $\kappa$ parameters should be positive independently.
Then, the positivity of the temperature $T=\frac{a}{2\pi}$ is
guaranteed by these two physical conditions. This in turn means
that the $f(r_h)=0$ equation either has one solution located at
$r_h=\frac{a-\sqrt{a^2+4b}}{-2b}$ for $b>0$, or two solutions
located at $r_h^{\pm}=\frac{a\pm\sqrt{a^2+4b}}{-2b}$ for $b<0$
with the condition of $a^2\geq |4b|$ for these solutions to be
real. The above results give us the corresponding Euler equation
as $M=TS$. Also, it may worth to mention that in a classical point
of view, where there is no mass loss (no evaporation) for the
black hole, the positivity of the thermodynamic temperature as
$T=\frac{\partial M}{\partial S}$ can be understood from the
second law of thermodynamics. But including the mass loss due to
the BH evaporation, and also demanding the second law of
thermodynamics, the temperature by this definition will be
negative. A rather similar situation happens when the Hawking
horizon temperature is defined by the surface gravity, i.e
$T_H=\frac{K}{2\pi}$, which represents the gravitational
acceleration as measured by the asymptotic observer for an
infalling object to the BH. Positivity of temperature then means
that gravity force is attractive and vice versa. However, for the
inner horizon of Reissner-Nordstrom BH, one finds a negative
surface gravity, representing a gravitational repulsion effect,
and then a  negative temperature. Thus, one may consider only the
absolute values for both these definitions of temperature.

Besides of what we obtained till now for the special case of
$b=\frac{\kappa\beta}{3}$, in which $\kappa\beta$ plays the role
of cosmological constant, one can write the total solution of the
field equations~(\ref{gr}) as

\begin{eqnarray}\label{met3}
&&\rho(r)=-p_r(r)=\frac{1}{\kappa}(\frac{2a}{r}+3b-\kappa\beta),\nonumber\\
&&p_t(r)=-\frac{1}{\kappa}(\frac{a}{r}+3b-\kappa\beta),
\end{eqnarray}

\noindent which clearly shows that at very large distances, we
face a cosmological constant-like source. Again, the obtained
solution~(\ref{met3}) represents that the fluid supporting the
geometry is anisotropic in general. Interestingly, in the
asymptotic region, i.e for $r\rightarrow\infty$, the fluid tends
to be isotropic with
$\rho(r)=-p_r(r)=-p_t(r)=\frac{3b}{\kappa}-\beta$. Then, for the
asymptotic region, $\frac{3b}{\kappa}-\beta$ plays the role of
cosmological constant. The latter can again establish a relation
between the values of $\beta$, $\kappa$, $b$ and the cosmological
constant $\Lambda$ (the current value of the dark energy density)
as

\begin{eqnarray}\label{met4}
&&\Lambda=3b-\kappa\beta,
\end{eqnarray}

\noindent meaning that both $b$ and $\beta$ may contribute in
forming $\Lambda$. Hence, if $b=\frac{\kappa\beta}{3}$ then
$\Lambda=0$ which is in agreement with the previous solution where
in the (anti)-de Sitter sector of metric ($b\neq0$) is cancelled
in the field equations~(\ref{gr1}) by adopting the
$b=\frac{\kappa\beta}{3}$ condition. We also see that, unlike the
previous solution for which the $b=0$ case does not cover the
(anti)-de Sitter geometry,  in the present solution even if we set
$b=0$, $\beta$ can take on the role of cosmological constant.
Eq.~(\ref{met4}) also implies that both the $\frac{b}{\kappa}$ and
$\beta$ terms have the same dimension, a result compatible with
the outcome of previous solution based on the
$b=\frac{\kappa\beta}{3}$ case.

In this situation, the mass circumscribed by the horizon is
obtained as

\begin{eqnarray}\label{mass2}
m(r_h)\equiv \mathcal{M}=\frac{a}{\kappa}A+\frac{\Lambda}{\kappa}
V,
\end{eqnarray}

\noindent where $V=\frac{4\pi}{3}r_h^3$. Combining this result
with the $S=\frac{2\pi}{\kappa}A$ relation \cite{mv}, and bearing
the thermodynamic definitions of temperature ($T$) and pressure
($P$) in mind, one reaches at

\begin{eqnarray}\label{Temp1}
&&T=\frac{\partial \mathcal{M}}{\partial S}\big|_{V=constant}=\frac{a}{2\pi},\nonumber\\
&&P=-\frac{\partial \mathcal{M}}{\partial
V}\big|_{S=constant}=-\frac{\Lambda}{\kappa}.
\end{eqnarray}
It finally leads to the Euler equation
$\mathcal{M}=TS-PV$. The Hawking temperature corresponding to
metric~(\ref{met1}) is also evaluated as \cite{pois,11}

\begin{eqnarray}\label{haw}
T_h=\frac{|\frac{df(r)}{dr}|_{r=r_h}}{2\pi}=\frac{\sqrt{a^2+4b}}{2\pi},
\end{eqnarray}
which indicates $T=T_h$ only if $b=0$. 
Therefore, if we are looking for the solutions with the same
Hawking and thermodynamic temperatures, then we have only one
possibility as $b=0$ and $a>0$. Regarding (\ref{haw}), the
condition of $a^2\geq |4b|$ for $b<0$ case is also required here
for the temperature $T_h$ to be real, similar to what we
discussed after Eq.~(\ref{mass20}).

Now, let us consider the generalized Misner-Sharp mass ($M_{MS}$)
confined to the horizon $r_h$ of metric~(\ref{met}) \cite{mv}

\begin{eqnarray}\label{misner}
&&M_{MS}=\frac{4\pi}{\kappa}\bigg[r_h+\kappa\bigg(\int\lambda[\frac{d(r^2f^{\prime}(r))}{dr}-2(1-\frac{d(rf(r))}{dr})]dr\bigg)_{r=r_h}\bigg].
\end{eqnarray}

\noindent Here, prime denotes the derivative with respect to time.
Using the $\lambda=\frac{\beta}{R}$ relation, in which
$R=-\frac{r^2f^{\prime\prime}(r)+4rf^{\prime}(r)-2[1-f(r)]}{r^2}$,
one can reach

\begin{eqnarray}\label{ms1}
M_{MS}=\frac{4\pi}{\kappa}\big[r_h-\frac{\kappa\beta}{3}r^3_h\big]=\frac{S}{2\pi
r_h}-\beta V,
\end{eqnarray}

\noindent where the $S=\frac{2\pi}{\kappa}A$ relation \cite{mv}
has been used to obtain the last equality. For the thermodynamic
temperature and pressure corresponding to~(\ref{ms1}), one finds

\begin{eqnarray}\label{ms2}
&&T_{MS}=\frac{\partial M_{MS}}{\partial S}\big|_{V=constant}=\frac{1}{2\pi r_h},\nonumber\\
&&P_{MS}=-\frac{\partial M_{MS}}{\partial
V}\big|_{S=constant}=\beta,
\end{eqnarray}

\noindent which again indicates that whenever the $M_{MS}$ mass is
considered, $\kappa\beta$ plays the role of cosmological constant.
Also, we have $T_{MS}=T_h$ only if $a=0$, which is nothing but the
(anti)-de Sitter spacetime for which the Cai-Kim temperature
($\frac{1}{2\pi r_h}$) is equal to the Hawking temperature
\cite{Caikim}. Therefore, the above results suggest that the
effects of $b$ and $a$ are stored in $r_h$, and the effect of
$\beta$ is appeared directly as the coefficient of volume (the
thermodynamic pressure). At this step, one can realize another
difference between the Einstein theory and our studied version of
Rastall theory. Considering the vacuum case of EECC
($G_{\mu\nu}=-\Lambda g_{\mu\nu}$), one reaches at (anti)-de
Sitter solution for (\ref{met}) in which its Misner-Sharp mass is
given by $\frac{r_h}{2G}$. This clearly differs from the $a=0$
limit (or equally, the (anti)-de Sitter limit of~(\ref{met1})) of
Eq.~(\ref{ms1}). Thus, the effect of cosmological constant is
stored only in $r_h$, and this is another difference between
Eq.~(\ref{gr1}) and EECC. Finally, it is worthwhile to mention
that $P_{MS}=P$ only if $b=0$ leading to $T_h=T\neq T_{MS}=0$.

\subsection{Conformally flat BHs in the Rastall theory}

For this theory, the horizon entropy meets the relation
$S=\frac{2\pi}{\kappa_R}A$ and the Newtonian limit leads to
\cite{ms}

\begin{eqnarray}\label{gamma1}
\kappa_R=\frac{4\gamma-1}{6\gamma-1},\ \
\xi=\frac{\gamma(6\gamma-1)}{4\gamma-1},
\end{eqnarray}

\noindent in which $\gamma=\kappa_R\xi$, and we assumed $8\pi G=1$.
In this subsection, we use the index $R$ to indicate that we work
in the original Rastall framework. Inserting the conformally flat
solution~(\ref{met1}) into the field equations~(\ref{r1}) and by
using~(\ref{gamma1}), one reaches

\begin{eqnarray}\label{r2}
&&\rho(r)=-p_r(r)=\frac{6\gamma-1}{4\gamma-1}\big[\frac{2a(1-3\gamma)}{r}+3b(1-4\gamma)\big],\\
\nonumber
&&p_t(r)=\frac{6\gamma-1}{4\gamma-1}\big[\frac{a(6\gamma-1)}{r}+3b(4\gamma-1)\big],
\end{eqnarray}

\noindent addressing us an anisotropic fluid. Now, following the
recipe led to Eqs.~(\ref{mass2}) and~(\ref{Temp1}), we obtain

\begin{eqnarray}\label{r3}
&&m(r_h)\equiv
\mathcal{M}_R=\frac{a(1-3\gamma)(6\gamma-1)}{4\gamma-1}A-3b(6\gamma-1)
V,\nonumber\\ &&T_R=\frac{\partial \mathcal{M}_R}{\partial S}\big|_{V=constant}=\frac{a(1-3\gamma)}{2\pi},\nonumber\\
&&P_R=-\frac{\partial \mathcal{M}_R}{\partial
V}\big|_{S=constant}=3b(6\gamma-1).
\end{eqnarray}

\noindent Eqs.~(\ref{r2}) and~(\ref{r3}) represent that in the
present framework, the $3b(1-6\gamma)$ term denotes the
thermodynamics pressure corresponding to the cosmological constant
$3b(4\gamma-1)$. In the Rastall framework, by considering the
generalized Misner-Sharp mass ($\mathcal{M}_R^{MS}$) confined to
radius $r_h$, expressed as \cite{ms}

\begin{eqnarray}\label{misner2}
\mathcal{M}_R^{MS}=\frac{6\gamma-1}{2(4\gamma-1)}[(1-2\gamma)r_h+\gamma
r_h^2 f^{\prime}(r_h)],
\end{eqnarray}

\noindent and since $S=\frac{2\pi}{\kappa_R}A$ \cite{ms}, one can
get

\begin{eqnarray}\label{r4}
&&\mathcal{M}_R^{MS}=\frac{1-\gamma(2+ar_h)}{2\pi r_h}S-6b\lambda
V,\nonumber\\ &&T_R^{MS}=\frac{\partial \mathcal{M}_R^{MS}}{\partial S}\big|_{V=constant}=\frac{1-\gamma(2+ar_h)}{2\pi r_h},\nonumber\\
&&P_R^{MS}=-\frac{\partial \mathcal{M}_R^{MS}}{\partial
V}\big|_{S=constant}=6b\lambda.
\end{eqnarray}

In summary, our results (Eqs.~(11-12),~(14),~(17) and
Eqs.~(20-22)) show that in both the Rastall theory and its special
generalized case, the thermodynamic pressure obtained by accepting
the mass definition~(\ref{mass0}) is equal to the pressure of the
cosmological constant candidate appeared in the field equations, a
result which is not obtained by employing the generalized
Misner-Sharp mass. In addition, the obtained thermodynamic
temperatures are not always equal to the Hawking temperature.

\section{Non-singular Black hole solutions}
\subsection{Non-singular BHs in the generalized Rastall theory}

It is important recalling that in GR, as well as in extended
theories of gravity (including the particular subclass of the
Rastall theory which is the framework of this paper) an unsolved
problem concerning BHs is the presence of a space-time singularity
in their core. Such a problem was present starting from the first
historical papers concerning BHs
\cite{Schwarzschild,Birkhoff,Chandrasekhar} and was generalized in
the famous paper by Penrose \cite{Penrose}. It is a common belief
that this problem can be solved when a correct quantum gravity
theory is obtained. For the sake of completeness, it is important
recalling some issues which dominate the question about the
existence or non-existence of BH horizons and singularities from
both of the theoretical and observational points of view, and
proposing some ways to remove BH singularities also at a classical
level, i.e. without taking into account of a quantum gravity
theory. Interesting alternatives to singular BHs are the so called
\textit{eternally collapsing objects} (ECO),
\textit{magnetospheric eternally collapsing objects}, (MECO) and
\textit{nonlinear electrodynamics} (NLED) \textit{objects}. An ECO
is a gravitationally compact mass supported against gravity by an
internal radiation pressure \cite{Mitra}. In its outer layers of
mass, a plasma with some baryonic content is supported by a net
outward flux of momentum via radiation \cite{Mitra}. Concerning
MECOS, one can postulate that some physical reason as, why the
existence of magnetic field should prevent formation of any event
horizon, can emerge by considering contracting plasma which is
threaded by a self-magnetic flux \cite{Robertson,Leiter,Schild}.
It is a good approximation to assume that the flux remains
conserved. Even if it is not conserved, there should be a finite
flux all the way \cite{Robertson,Leiter,Schild}. Let us presume
that the plasma ball collapses inside its event horizon. The
region inside the event horizon is trapped, and this means, no
lines of force can emanate out of the plasma ball, then a local
observer siting at a radius larger than the Schwarzschild radius
will not see any magnetic field \cite{Robertson,Leiter,Schild}.
Indeed by no hair theorem, a neutral BH has no magnetic field.
This means, the entire magnetic flux must vanish before the plasma
ball can enter its event horizon \cite{Robertson,Leiter,Schild}.
Conversely, a plasma ball endowed with initial magnetic field
cannot become a BH unless it can destroy its entire magnetic
field. But why should the ball destroy its entire magnetic field
to enter the event horizon, which in the BH folk lore, is a mere
coordinate singularity which a comoving observer cannot notice at
all. Hence, a plasma with initial magnetic field cannot form a BH
\cite{Robertson,Leiter,Schild}. This is consistent with a paper by
the Nobel Laureate K. Thorne, who showed back in 1965 that pure
magnetic energy would not collapse into a BH state \cite{Thorne}.
On the other, it has been recently shown that NLED objects can
remove BH singularities too. In Ref \cite{Corda}, a particular
solution of Einstein field equation for a model of star supported
against self-gravity entirely by radiation pressure has been
discussed. In such a solution trapped surfaces as defined in GR
are not formed during a gravitational collapse, and hence the
singularity theorem on black holes, as proposed in \cite{Penrose},
cannot be applied. More in general, NLED effects turn out to be
important as regard to the mass-radius relation \cite{Corda},
which is maximum for a BH.

Also, there is another interesting proposal which concerns the
possibility to replace the Schwarzschild singularity with a de
Sitter vacuum. This proposal started from an idea by Sakharov
\cite{Sakharov}, who considered a negative density in the equation
of state for super-high density, and by Gliner \cite{Gliner}, who
interpreted such a negative density as corresponding to a vacuum
and suggested that it could be the final state of the
gravitational collapse. This approach has been recently considered
in \cite{non}.

Let us follow Ref. \cite{non} and consider the energy density

\begin{eqnarray}\label{den1}
&&\rho(r)=a\exp(-\frac{r^3}{b^3}),
\end{eqnarray}

\noindent where $a$ and $b$ are unknown constants in general. Note
that $a$ and $b$ parameters here are different than in the
previous solution. In this manner, solving Eq.~(\ref{gr1}), one
reaches at

\begin{equation}\label{met5}
f(r)=1-\frac{C_1}{r}-\frac{\kappa\beta}{3}r^2+\frac{\kappa
ab^3}{3r}\exp(-\frac{r^3}{b^3}),
\end{equation}

\noindent and

\begin{equation}\label{jinguli}
p_r(r)=-\rho(r),\ p_t(r)=(\frac{3r^3}{2b^3}-a)\rho(r),
\end{equation}

\noindent which is an isotropic source. For this solution, the
Ricci square and Ricci scalar are well-behaved at the
$r\rightarrow0$ limit. Moreover, the Weyl and Riemann squares are
also well-behaved at the $r\rightarrow0$ limit whenever

\begin{eqnarray}\label{met6}
&&C_1=\frac{b^3}{3\kappa a},\ \kappa=\pm\frac{1}{a},
\end{eqnarray}

\noindent where regarding (\ref{met5}), $C_1$ appears as the mass
parameter. Here, the positivity of both the $a$ and $b$ parameters
are guaranteed by the weak energy condition and by the positivity
of the mass parameter $C_1$. Moreover, bearing the
$S=\frac{2\pi}{\kappa}A$ \cite{mv} relation in mind and noting
that entropy should be positive, the possibility of
$\kappa=-\frac{1}{a}$ is ruled out.

To show that the solution (\ref{met5}) with constraints
$C_1=\frac{b^3}{3}$ and $\kappa=\frac{1}{a}$~(\ref{met6}) is
regular everywhere, one can use the following
Eddington-Finkelstein coordinate transformation

\begin{equation}
du=dt+\frac{dr}{1-\frac{\beta}{3a}r^2-\frac{b^3}{3r}\big(1-\exp(-\frac{r^3}{b^3})\big)},
\end{equation}

\noindent to write the metric (\ref{met5}) as

\begin{equation}\label{mtg}
ds^2 =-\left(1-\frac{\beta}{3a}r^2-\frac{b^3}{3r}\big(1-\exp(-\frac{r^3}{b^3})\big)\right)du^2
+2dudr+r^2d\Omega^2,
\end{equation}

\noindent representing that the solution is non-singular
everywhere (in both $r=0$ and $r=r_h$). One notes that due to the
existence of the non-minimal coupling, the asymptotic nature of
the solutions (\ref{met5}) (or (\ref{mtg}))  is different than
\cite{non} for large $r$ values which coincides with the
Schwarzschild solution and for small $r$ values behaves like the
de Sitter solution. The solution (\ref{met5}) has the same
internal nature as in \cite{non} due to the considered specific
vacuum stress-energy momentum tensor in (\ref{den1}) and
(\ref{jinguli}), but possesses a de Sitter asymptote, instead of
Schwarzschild, with a cosmological term $\frac{\kappa\beta}{3}$
induced by the non-minimal coupling property of the background
theory.  In this manner,
 $f(r_h)=0$ has a solution at
$r_h=b$ if

\begin{eqnarray}\label{met7}
\beta=a(\frac{3-b^2(1-e^{-1})}{b^2}).
\end{eqnarray}

\noindent This is a relation between the non-minimal coupling
parameter $\beta$ (or equally $\lambda$) and the BH's properties
(or equally the energy source). In both sides of $r_h=b$ horizon,
we have $f(r<b)>0$ and $f(r>b)<0$. This means that the metric
changes its signature from $(-,+,+,+)$ to $(+,-,+,+)$ at $r=b$.

Now, without considering $\kappa=\frac{1}{a}$, using
Eq.~(\ref{den1}) and the area-entropy relation
$S=\frac{2\pi}{\kappa}A$ \cite{mv}, one reaches

\begin{eqnarray}\label{mass3}
m(b)\equiv \tilde{M}=4\pi\int_{0}^{b}\rho(r)r^2dr=\frac{\kappa
ba}{6\pi}S-\frac{a}{e}V,
\end{eqnarray}

\noindent as the BH's mass confined to the horizon located at
$r_h=b$ where $V=\frac{4\pi}{3}b^3$ and $A=4\pi b^2$. The
thermodynamic temperature and pressure can be obtained as

\begin{eqnarray}\label{Temp2}
&&T=\frac{\partial \tilde{M}}{\partial S}\big|_{V=constant}=\frac{\kappa ab}{6\pi},\nonumber\\
&&P=-\frac{\partial \tilde{M}}{\partial
V}\big|_{S=constant}=\frac{a}{e},
\end{eqnarray}

\noindent respectively. We see that the temperature of the system
is positive only if $\kappa=\frac{1}{a}$ which leads to
$T=\frac{b}{6\pi}$, and hence, the Euler equation takes the form
of $\tilde{M}=TS-PV$. For this solution, we also find
$T_h=\frac{b|1-\frac{6}{b^2}-4e^{-1}|}{6\pi}$ as the Hawking
temperature which is equal to the thermodynamic
temperature~(\ref{Temp2}) only if $1-\frac{6}{b^2}-4e^{-1}=\pm1$.
This constraint yields $b^2=\frac{3}{1-2e^{-1}}$. Eq.~(\ref{met7})
has also an interesting solution for $f(r_h)=0$ located at
$b=\frac{3}{1-e^{-1}}$ for which $\beta$, and thus $\lambda$, will
be zero. In this situation, the generalized Rastall theory reduces
to the Einstein theory ($\kappa=8\pi G$ \cite{mv}) in the absence
of cosmological constant. Hence, Eq.~(\ref{met6}) implies
$\kappa=\frac{1}{a}=8\pi G$ whenever $\beta=0$ which is in
agreement with Ref. \cite{non}.

Bearing Eq.~(\ref{misner}) in mind, one can easily see
Eq.~(\ref{ms1}) and thus Eq.~(\ref{ms2}) are still valid. It is
due to this fact that these results are independent of $f(r)$, a
property which comes from the $\beta=\lambda R$ constraint. Of
course, it should be noted that, for the above solution, $r_h$
differs from that of the conformal BHs, and it is obtained by
finding out the zeros of Eq.~(\ref{met5}). It is also easy to
check Eq.~(\ref{met5}) is valid for EECC if we change
$\kappa\beta$ and $\kappa$ as $\Lambda$ and $8\pi G$,
respectively. Hence, since the Misner-Sharp mass of the Einstein
framework takes the $\frac{r_h}{2G}$ form, once again, we can see
that the Misner-Sharp mass of the obtained solution is different
in the frameworks of the considered generalized Rastall theory and
EECC.

\subsection{Non-singular BHs in the Rastall theory}

Here, we do not focus on finding out the non-singular BHs
solutions, corresponding to Eq.~(\ref{den1}), in the Rastall
framework, due to the complexity of the field equations. This
issue requires further investigations that the results of which
will be reported as an independent research work. It is also
useful to mention that the Gaussian BHs, in which
$\rho\sim\exp(-r^2)$, have previously been studied
\cite{nonras1,nonras}.

\section{Conclusion}

Focusing on a special subclass of the generalized Rastall theory
\cite{genras}, whose field equations are mathematically similar to
EECC, as well as on the original Rastall theory, we show that the
ability of spacetime to couple with the matter fields in a
non-minimal way can either play the role of cosmological constant,
see~(\ref{gr1}), or affect it, see (\ref{met4}) and~(\ref{r2}),
depending on the energy source supporting the geometry. The
conformally flat BHs have also been derived, and some of their
properties were studied in both frameworks. Additionally and only
in the framework of field equations~(\ref{gr1}), a non-singular BH
has been obtained. Relations between the parameters of obtained
non-singular BH, those of the energy source and the non-minimal
coupling were also addressed. We found out that the ability of
spacetime to non-minimally couple with the matter fields can
affect the location of the horizons of spacetime even whenever OCL
is valid ($\beta=\textmd{constant}\neq0$).

The Euler equations corresponding to the solutions have also been
derived. It is shown that for the metric~(\ref{met}), if the
integral~(\ref{mass0}) and $S=\frac{2\pi}{\kappa}A$ are accepted
as the mass \cite{non} and entropy \cite{mv} of BH, respectively,
then the thermodynamic pressure and temperature, satisfying the
Euler equation, are not always equal to the pressure components
obtained using the field equations, and also the Hawking
temperature, respectively, a result also valid in the Einstein
theory (the $\beta=0$ and $\kappa=8\pi G$ limits of the obtained
relations) \cite{thermotbh}. The requirements of having solutions
with the same thermodynamic and Hawking temperatures have also
been studied.

The consequences of considering the generalized Misner-Sharp mass
have also been investigated, and the corresponding Euler equations
have been derived in both frameworks. It seems that the
thermodynamic pressure obtained by accepting the mass
definition~(\ref{mass0}) is equal to the pressure of the
cosmological constant candidate appeared in the field equations
compared with the situation in which the corresponding generalized
Misner-Sharp mass is employed. In this situation, the obtained
thermodynamic temperature is not always equal to the Hawking
temperature.

\section*{Acknowledgments}
We are so grateful to the anonymes referee for valuable comments
and suggestions. 

\end{document}